\documentclass[
reprint,
superscriptaddress,
amsmath,amssymb,
aps,
prl,
]{revtex4-2}

\usepackage{graphicx}% Include figure files

\usepackage{dcolumn}% Align table columns on decimal point
\usepackage{enumitem}
\usepackage{bm}% bold math
\usepackage[colorlinks=true, linkcolor=blue, citecolor=blue, urlcolor=blue]{hyperref}

\begin{document}

\preprint{APS/123-QED}
%TC:ignore
\title{The Nature of Turbulence at Sub-Electron Scales in the Solar Wind}% Force line breaks with \\

\author{Shiladittya Mondal}
\email{shiladittya.mondal@qmul.ac.uk}
\author{Christopher H. K. Chen}%
\author{Davide Manzini}%
\affiliation{%
 Department of Physics \& Astronomy, Queen Mary University of London, London E1 4NS, United Kingdom
}%

\date{\today}% It is always \today, today,

\begin{abstract}

The nature of turbulence at sub-electron scales has remained an open question, central to understanding how electrons are heated in the solar wind. 
This is primarily because spacecraft measurements have been limited to magnetic field fluctuations alone. We resolve this by deriving new high-resolution density fluctuations from spacecraft potential measurements of Parker Solar Probe resolving scales smaller than the electron gyro-radius ($\rho_e$). A systematic comparison of the density and magnetic spectra shows that both steepen near the electron scales. Notably, the density spectrum exhibits slopes close to $-10/3$, while the magnetic spectrum becomes consistently steeper than the density spectrum at scales smaller than $\rho_e$, indicating that the turbulence becomes electrostatic. These results are consistent with theoretical predictions of an electron entropy cascade, which may explain the irreversible dissipation of turbulent energy at sub-$\rho_e$ scales. The magnetic spectrum, however, is not as steep as expected for the electron entropy cascade, which may be due to limited signal-to-noise ratio and the presence of weakly damped electromagnetic fluctuations near $\rho_e$.
\end{abstract}

\maketitle
%TC:endignore

\noindent
\paragraph*{Introduction:}
{
%%%%%%%%%%%%%%%%%%% PARAGRAPH 1 %%%%%%%%%%%%%%%%%%%%

Turbulence in the solar wind drives a nonlinear cascade of energy injected at large scales to smaller kinetic scales, where it gets dissipated \cite{Frisch_1995, Tu_1995, Biskamp_2003, Bruno2013}.
This is well established from observations of magnetic power spectra, which exhibit a broadband wavenumber power-law scaling $P(k)\sim k^{\alpha}$ with $\alpha\simeq-3/2$ to $-5/3$ at magneto-hydrodynamic (MHD) scales \cite{Chen_2020, Wilson_2021, sioulas_2024, Mondal_2025}, consistent with models of Alfvénic turbulence \cite{Chen_2016, Schekochihin_2022, Sasmal_2025, Alberti_2025}.
At sub-ion scales, between the ion gyro-radius ($\rho_i$) and electron inertial scale ($d_e$), density ($n$) and magnetic field ($B$) spectra have been extensively measured, showing slopes close to $-7/3$ or $-8/3$ \cite{Chen_2013, Roberts_2018}, in agreement with predictions of kinetic Alfvén (KA) turbulence \cite{Howes_2006, Cho_2009, Boldyrev_2012, Boldyrev_2013, Howes_2011, Schekochihin_2009, Vincent_2024}.
In the range between $d_e$ and $\rho_e$, an increase in magnetic compressibility was found, with the magnetic spectrum showing a slope near $-11/3$, interpreted as a transition from KA to inertial-KA cascade under low electron beta ($\beta_e$) conditions, in the magnetosheath \cite{Chen2017}.
In contrast, instrumental limitations have prevented a similar extent of progress at sub-electron scales (below the electron gyro-radius, $\rho_e$), with measurements affected by low signal-to-noise ratio (SNR) and restricted to magnetic fluctuations alone. As a result, the nature of turbulence at these scales has remained an open question.

%%%%%%%%%%%%%%%%%%% PARAGRAPH 2 %%%%%%%%%%%%%%%%%%%%

At sub-electron scales, a few studies have reported an exponential roll-off of the magnetic spectrum, interpreted as fluid-like dissipation \cite{Alexandrova_2009, Alexandrova_2012, Alexandrova_2021}. Others, however, reported a steeper power-law spectrum ($\alpha \simeq -4$), attributed to enhanced Landau damping \cite{Sahraoui_2009, Kiyani_2009, Sahraoui_2013}. Even steeper slopes ($\alpha\simeq -5$) have been found in the Earth's magnetosheath, where magnetic field measurements had higher signal-to-noise ratio (SNR) than in the solar wind \cite{Huang_2014}.
Given these very limited observations and an incomplete understanding of turbulence in this range, several theoretical and numerical frameworks have been proposed.
To explain the observed steeper magnetic spectrum, an incompressible electron-MHD model was suggested, predicting a slope of $-11/3$ at $k_\perp d_e \gg 1$ \cite{Meyrand_2010}, although it neglects damping effects that become important at these scales. Here, $k_\perp$ is the wave vector perpendicular to the local mean magnetic field. Simulations of whistler-driven turbulence have also been shown to produce steeper slopes $\lesssim -4.5$ below $d_e$ \cite{Chang_2011, Gary_2012}.
Kinetic simulations incorporating effects such as Landau damping likewise tend to produce either exponential or steep power-law spectra \cite{Camporeale_2011, TenBarge_2013, Parashar_2018, Arro_2022}.
However, since these formalisms do not account for density fluctuations, it becomes difficult to distinguish between them using only magnetic field observations.

%%%%%%%%%%%%%%%%%%% PARAGRAPH 3 %%%%%%%%%%%%%%%%%%%%

Finally, a phase-space cascade known as the electron entropy cascade is proposed for $k_\perp \rho_e\gg1$ \cite{Schekochihin_2008, Schekochihin_2009}.
This cascade arises from nonlinear interactions between electrostatic fluctuations and the perturbed electron distribution function, generating fine-scale structure in both velocity and position space.
This is expected to cause finite entropy production and irreversible electron heating.
In position space, the cascade is predicted to have spectral scalings: $E_n \sim k_\perp^{-10/3}$ and $E_B \sim k_\perp^{-16/3}$ \cite{Schekochihin_2008, Schekochihin_2009}.
Similar scaling laws apply to the ion entropy cascade, for which the shallower density scaling has thus far only been observed in laboratory experiments \cite{Kawamori_2013} and gyro-kinetic simulations \cite{Tatsuno_2009, Navarro_2011}. However, it has not yet been detected for either ions or electrons in space or astrophysical plasmas.
A spectrum of ion velocity-space structures has been measured, interpreted as an ion phase-space cascade \cite{Servidio_2017},
however, such analysis is not possible at sub-electron scales with the current resolution of particle instruments.
Consequently, the physics at these scales has thus far remained poorly understood. This motivates a study of the density fluctuations around electron scales to determine the nature of turbulence and distinguish the underlying physics.

%%%%%%%%%%%%%%%%%%% PARAGRAPH 4 %%%%%%%%%%%%%%%%%%%%

In this Letter, we present new high-resolution density measurements from spacecraft potential data of Parker Solar Probe (PSP) \cite{Fox_2016} across electron scales. By comparing density and magnetic spectra, we show that both steepen below $\rho_e$ but exhibit distinct slopes, with the magnetic spectrum being steeper. These observations indicate a transition to electrostatic turbulence and are consistent with the predictions of an electron entropy cascade, which may explain irreversible dissipation of turbulent energy at sub-electron scales.
}

\noindent
\paragraph*{Data \& Measurement techniques:}{

%%%%%%%%%%%%%%%%%%% PARAGRAPH 5 %%%%%%%%%%%%%%%%%%%%

We use high-resolution search coil magnetometer (SCM) and antenna potential burst mode data from the FIELDS instrument suite on PSP, sampled at $\sim 19\,\mathrm{kHz}$ \cite{Bale_2016, Malaspina_2016, Jannet_2021}.
Data from encounters E10–E18 were surveyed for intervals with enhanced fluctuation amplitudes near the Sun, where density and magnetic spectra could be resolved at scales below $\rho_e$.
Although $\rho_e$ and $d_e$ shift beyond instrumental resolution near perihelion, several $27\,\mathrm{s}$ intervals within $0.15–0.2\,\textrm{au}$ were identified where sub-$\rho_e$ scales could be resolved.
We selected 76 such intervals without dust hits or wave activity, exhibiting $\textrm{SNR}\gtrsim 10$ at the observed spectral break around $d_e$ (detailed in the next section). 
The intervals used do not introduce any systematic biases and constitute a representative sample spanning a range of solar wind conditions near the Sun, with plasma beta $\beta_i \simeq \beta_e \simeq 0.4$ and a temperature ratio $T_i/T_e \simeq 1$.

%%%%%%%%%%%%%%%%%%% PARAGRAPH 6 %%%%%%%%%%%%%%%%%%%%

Since SPAN, SPC \cite{Kasper2016}, and QTN \cite{Moncuquet_2020} density measurements have a low sampling rate ($\sim0.14\,\mathrm{Hz}$), we use spacecraft potential ($V_\textrm{sc}$) as a proxy for density to resolve scales smaller than $\rho_e$.
This method is based on the following principle: Solar illumination charges the spacecraft body positive via photoemission, attracting nearby electrons and thereby making $V_\mathrm{sc}$ sensitive to local electron density ($n$) \cite{Pedersen_1995, Kellogg_2005, Chen_2012a, Khotyaintsev_2021, Mozer_2022}.
The resulting balance between photoelectric and thermal electron currents leads to an exponential relation:
$n \propto e^{- a V_\mathrm{sc}/V_\mathrm{pe}+C_\mathrm{b}}$,
where $V_\mathrm{pe}$ is the photoelectron thermal energy.
$V_\mathrm{sc}$ is computed as the negative average of the four PSP antenna potentials, $V_\mathrm{sc}=-(V_1+V_2+V_3+V_4)/4$, which includes the antenna bias voltage. 
To account for this bias, an offset parameter $(C_\mathrm{b})$ was included.
Other currents, such as ion or secondary electron currents, may be present but are negligible.
This relation is indeed observed between electron QTN density measurements and $V_\mathrm{sc}$ for the intervals used. Typically at sub-ion scales, $\delta V_\textrm{sc}/V_{\mathrm{pe}}\ll 1$, implying $\delta n \propto \delta V_\mathrm{sc}$, up to first order. Thus, the two spectra share the same shape and the $V_\textrm{sc}$ spectrum can be used to measure the density spectral slopes.

%%%%%%%%%%%%%%%%%%% PARAGRAPH 7 %%%%%%%%%%%%%%%%%%%%

It must be noted that $V_\mathrm{sc}$ does not respond instantaneously to density variations, but adjusts exponentially with a time constant $\tau_c \approx C V_\mathrm{pe} / I_{\rm th,e}$ \cite{Chen_2013b}, where $C$ is the spacecraft capacitance and $I_{\rm th,e}$ the thermal electron current. For frequencies greater than or comparable to $\tau_c^{-1}$, the $V_\mathrm{sc}$ spectrum would show artificial steepening due to reduced frequency response.
In the near-Sun solar wind intervals used here, $\tau_c^{-1}\gtrsim 10\,\mathrm{kHz}$. Since our analysis is based on much lower frequencies ($\lesssim 1\,\mathrm{kHz}$), the simple linear dependence $\delta n \propto \delta V_{\mathrm{sc}}$ is well satisfied.

}

\noindent
\paragraph*{Results \& Discussion:}{

%%%%%%%%%%%%%%%%%%% PARAGRAPH 8 %%%%%%%%%%%%%%%%%%%%
\begin{figure}
    \centering
    \includegraphics[width=1\linewidth, trim=5 5 5 5, clip]{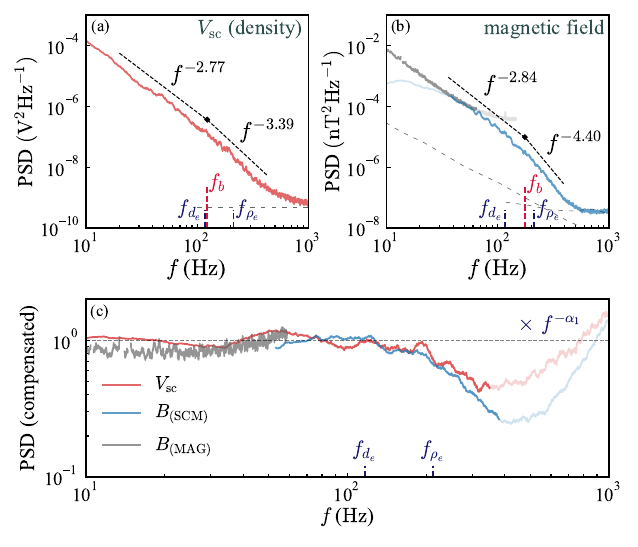}
    \caption{Power spectral density of (a) $V_\mathrm{sc}$ and (b) $B$ (MAG - dashed, SCM - solid). The least square fits (vertically shifted for clarity) and the break frequency ($f_b$) using Eq.~\ref{eqn1} are shown. Thin-dashed lines indicate the flat analog-to-digital converter and the frequency dependent SCM noise floors. (c) $V_\mathrm{sc}$ and $B$ spectra compensated by the sub-ion range scaling $(f^{-\alpha_1}P(f))$. The compensated spectra are shown as dotted lines for SNR $\lesssim 2$.}
    \label{fig1}
    \vspace{-2mm}
\end{figure}
For the selected intervals, the power spectral density (PSD) of $B$ and $V_\textrm{sc}$ were computed using fast Fourier transforms.
Fig.~\ref{fig1}(a,b) shows the $V_\textrm{sc}$ and trace $B$ spectra vs spacecraft frame frequency ($f$) for a typical interval on 2022-06-04, 20:54:10–20:54:37.
The trace $B$ spectrum is computed using only the $y$ and $z$ SCM components as the $x$ component is unavailable.
For this interval, the low-frequency $B$ spectrum using the flux-gate magnetometer (MAG) data is also shown (MAG data was rotated into SCM coordinates before computing the trace \cite{Bowen_2020}).
The spectra represented are smoothed using a \textcolor{black}{150 pt.} running mean window. Fig.~\ref{fig1}(c) shows the $V_\textrm{sc}$ and $B$ spectra compensated by the sub-ion range scaling.
The Doppler shifted frequencies corresponding to $k_\perp d_e=1$ and $k_\perp\rho_e=1$ are marked using, $f_{d_e} (f_{\rho_e})\simeq V_\mathrm{rl} \sin \theta_{BV_\mathrm{rl}}/2\pi d_e(\rho_e)$, assuming Taylor's hypothesis \cite{Taylor_1938}.
Here, $V_\mathrm{rl}$ is the magnitude of the relative velocity between the solar wind and the spacecraft, and $\theta_{BV_\mathrm{rl}}$ is the angle between $\mathbf{B}$ and $\mathbf{V_\mathrm{rl}}$. 
A piecewise power-law,
\begin{equation}
    P(f) =
    \begin{cases}
        A_1f^{\alpha_1}, & f < f_b \\
        A_2f^{\alpha_2}, & f \geq f_b
        \end{cases}
\label{eqn1}
\end{equation}
was fitted over the sub-ion to sub-electron scale range to each $V_\mathrm{sc}$ (hereafter $n$) and trace $B$ (SCM) spectrum using a least-square method. 

Fits were restricted to frequencies with SNR $\gtrsim2–3$. For $f \lesssim 35\,\mathrm{Hz}$, the flattening in $B$ spectra (faded solid blue line) caused by the reduced SCM frequency response, was also excluded. The fits provide the amplitude $A_1$, spectral slopes $\alpha_1$ and $\alpha_2$, and the break frequency $f_b$, with $A_2$ being determined from the other parameters.

\begin{figure}[t]
    \centering
    \includegraphics[width=1\linewidth, trim=5 5 5 5, clip]{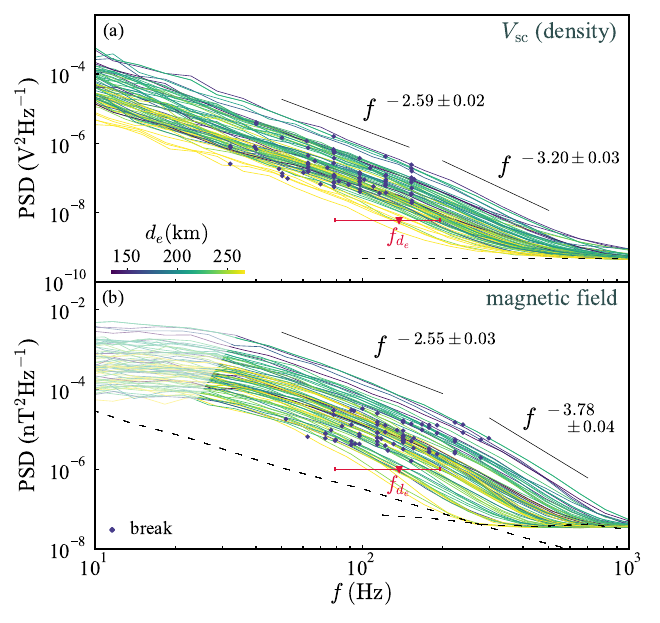}
    \caption{(a) $V_\mathrm{sc}$ and (b) $B$ spectra for all intervals smoothed using a variable width running mean window. Average slopes at sub-ion and sub-electron scales are shown. Blue diamonds mark the breaks obtained from fits. Mean $f_{d_e}$ and its variation across the intervals are shown by the red marker.}
    \vspace{-0.5cm}
    \label{fig2}
\end{figure}

%%%%%%%%%%%%%%%%%%% PARAGRAPH 9 %%%%%%%%%%%%%%%%%%%%

As depicted in Fig.~\ref{fig1}, the $n$ and $B$ spectra show similar slopes of $-2.77$ and $-2.84$, respectively, in the sub-ion  range ($f < f_{d_e}$).
A spectral break is identified from the fit at frequency $f_b$ approximately near $f_{d_e}$ or $f_{\rho_e}$ in both spectra.
Thereafter, the $B$ spectrum steepens to $-4.40$ while the $n$ spectrum is comparatively shallower at $-3.39$. This transition is clear from the compensated spectra, which steepen compared to the sub-ion range scaling ($f^{\alpha_1}$, horizontal dashed line) at $f_{d_e}$ or $f_{\rho_e}$.
A similar trend was observed across all studied intervals as evident from Fig.~\ref{fig2}.
This figure also marks the spectral breaks (blue diamonds), showing they appear approximately around $f_{d_e}$.
While a steepening in the $B$ spectrum has been reported before close to the electron scales, we show here that the $n$ spectrum also steepens, although to a lesser degree than the $B$ spectrum.
This difference is evident from the distribution of $n$ (Fig.~\ref{fig3}(a)) and $B$ (Fig.~\ref{fig3}(b)) spectral slopes obtained from the fits (Eq.~\ref{eqn1}) across all intervals.
In the sub-ion range ($f < f_b$), the distributions peak near $-8/3$, with mean slopes of $-2.59\pm0.02$ for $B$ and $-2.55\pm0.03$ for $n$, with the uncertainties indicating the standard error of the mean.
This is consistent with previous observations \cite{Chen_2013, Roberts_2018}, and the theoretical and numerical predictions of KA turbulence, which suggest a slope close to $-7/3$ or $-8/3$ for both fields \cite{Howes_2006, Cho_2009, Boldyrev_2012, Boldyrev_2013, Howes_2011, Schekochihin_2009, Vincent_2024}.
%\textcolor{red}{This result of the $V_\text{sc}$ and $B$ spectra exhibiting similar slopes at sub-ion scales also acts as physics consistency check indicating that $V_\text{sc}$ acts as a good proxy for probing $n$ fluctuations.}
In the sub-electron range ($f > f_b$), both spectra steepen, but differ in slope. While $n$ shows a mean slope of $-3.20\pm0.03$, $B$ steepens to $-3.78\pm0.04$, occasionally reaching up to $-4.8$. This is the main result of this Letter, with this difference in slope indicating a breakdown of KA turbulence close to $\rho_e$.

%%%%%%%%%%%%%%%%%%% PARAGRAPH 10 %%%%%%%%%%%%%%%%%%%%

However, before further interpreting the observations at sub-$\rho_e$ scales, we discuss the transition from sub-ion ($k_\perp d_e < 1$) to sub-electron ($k_\perp \rho_e > 1$) scales, which has been a subject of debate.
Previous studies suggested that the $B$ spectrum either shows a break at $\rho_e$ followed by a steeper power-law \cite{Sahraoui_2009, Sahraoui_2013, Huang_2014}, or undergoes a smoother transition to an exponential roll-off \cite{Alexandrova_2009, Alexandrova_2012, Alexandrova_2021}.
We emphasize here that, although we use a piecewise power law with a single break to characterize the two regimes separately, the transition involves multiple physical processes and is not defined by either a single break or an exponential curve.
\begin{figure}[t]
    \centering
    \includegraphics[width=1\linewidth, trim=5 5 5 5, clip]{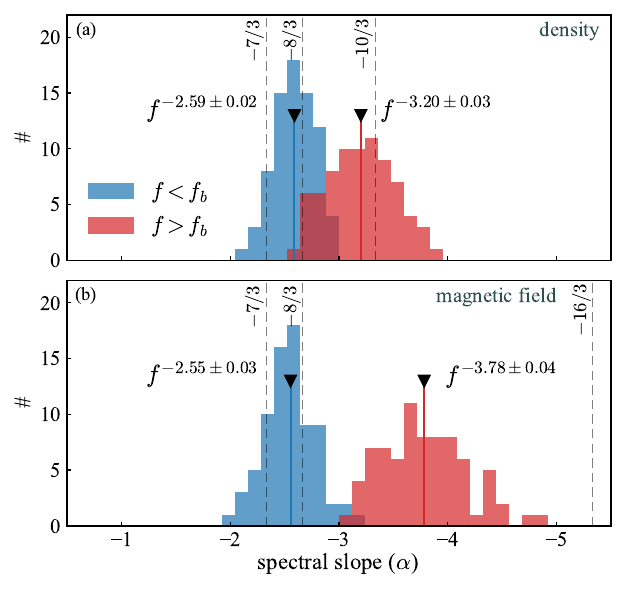}
    \caption{Distributions of the (a) $n$ and (b) $B$ spectral slopes obtained from the fits at sub-ion ($f<f_b$, in blue) and sub-electron ($f>f_b$, in red) scales. Vertical lines indicate the mean and standard error of the mean for each distribution.}
    \vspace{-5.5mm}
    \label{fig3}
\end{figure}
This can be understood from Fig.~\ref{fig4}. Panels (a) and (b) show the break positions identified by the fits in the $n$ and $B$ spectra ($f_{b,n}$ and $f_{b,B}$) relative to $f_{d_e}$ and $f_{\rho_e}$, respectively. Panel (c) shows the average trend of the $n$--$B$ spectral ratio across the intervals, normalized to the mean spectral power in the sub-ion range.
The breaks (dots) tend to occur around $d_e$ for both $n$ and $B$ spectra rather than $\rho_e$, although they were found to be weakly correlated with both scales. Their proximity to the $y=x$ dashed line indicates that both $n$ and $B$ spectra start to steepen at similar frequencies ($\sim f_{d_e}$), maintaining similar slopes until reaching $f_{\rho_e}$. This can be seen from the behavior of the normalized $n$--$B$ spectral ratio (bottom panel), which is constant throughout the sub-ion range and remains flat even near $f_{d_e}$.
For $f \gtrsim f_{\rho_e}$, the ratio increases, clearly indicating that the $B$ spectrum steepens more than the $n$ spectrum.
The steepening around $d_e$ may arise from enhanced electron Landau damping \cite{Chen_2019}, with the fluctuations still remaining Alfvénic in nature as $n$ and $B$ spectra exhibit similar slopes. A transition from KA to inertial-KA turbulence near $d_e$ may also be involved, for which both spectra are expected to steepen to a $-11/3$ slope \cite{Chen2017}.
Because of these different physical effects, the transition from sub-ion to sub-electron scales is likely not characterized by either a single well-defined break or an exponential curve.
\begin{figure}[t]
    \centering
    \includegraphics[width=1\linewidth, trim=5 5 5 5, clip]{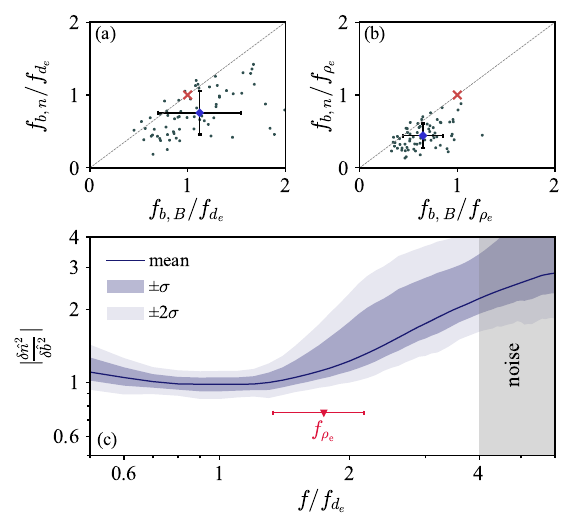}
    \caption{Break frequencies $f_{b,n}$ and $f_{b,B}$ obtained from fits relative to (a) $f_{d_e}$ and (b) $f_{\rho_e}$. (c) Mean trend of the normalized $n$–$B$ spectral ratio as a function of $f/f_{d_e}$. Violet shaded regions indicate the standard deviation ($\sigma$) across the intervals. The mean $f_{\rho_e}$ and its variation are shown by the red marker.}
    \vspace{-3mm}
    \label{fig4}
\end{figure}

%%%%%%%%%%%%%%%%%%% PARAGRAPH 11 %%%%%%%%%%%%%%%%%%%%

In the sub-electron range, again due to limited frequency resolution, it is difficult to distinguish whether the magnetic spectrum follows a power law or an exponential roll-off. However, the availability of density measurements allows an investigation of the physics in this range.
Fig.~\ref{fig3} shows that for $f>f_b$, the $n$ spectrum exhibits a mean slope of $-3.20 \pm 0.03$, while the $B$ spectrum is steeper with $-3.78 \pm 0.04$.
The steeper $B$ slope compared to $n$ provides evidence of turbulence becoming electrostatic as the electrons demagnetize at sub-$\rho_e$ scales. The density fluctuations, which are essentially the fluctuations in electrostatic potential show slopes close to the $-10/3$ prediction of the electron entropy cascade. Moreover, $B$ steepens relative to $n$ only for $f>f_{\rho_e}$ ($k_\perp \rho_e > 1$), as evident from Fig.~\ref{fig4}(c). These observations are consistent with the electron entropy cascade theory at sub-$\rho_e$ scales \cite{Schekochihin_2008, Schekochihin_2009}.
However, the $B$ spectrum does not reach slopes as steep as the theoretical $-16/3$ value.
To investigate whether this could be due to limited SNR, we use an empirical spectral model (Eq.~\ref{eqn2}), combining a continuous bi-power law with the instrument noise floor:

\begin{figure}[t]
    \centering
    \includegraphics[width=1\linewidth, trim=5 5 5 5, clip]{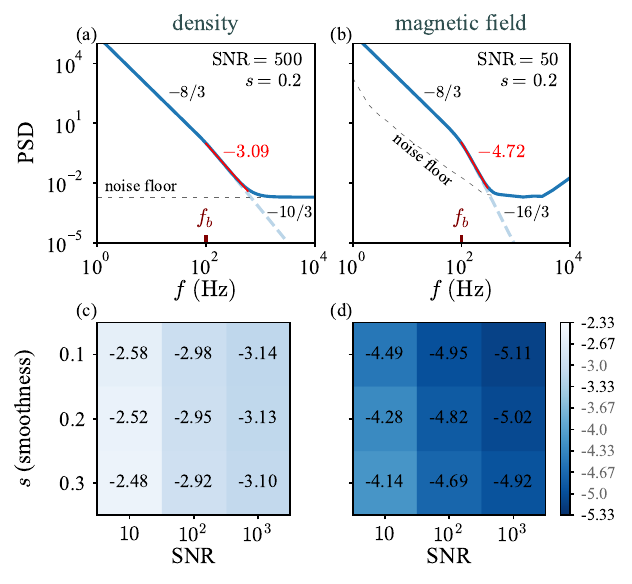}
    \caption{Empirical spectra of (a) $n$ and (b) $B$ from Eq.~\ref{eqn2}, with noise-free (dashed) and noise-added (solid) cases (units are arbitrary). Fits to the latter in the range $f_b < f < f_{\mathrm{SNR=2.5}}$ are marked in red. (c,d) Variation of slopes obtained from the fits to the empirical spectra of $n$ and $B$ with SNR and smoothness $(s)$.}
    \vspace{-5.5mm}
    \label{fig5}
\end{figure}

\begin{equation}
    B(f) = \left( \frac{f}{f_b} \right)^{\alpha_1} \left[ \frac{1}{2} \left( 1 + \left( \frac{f}{f_b} \right)^{1/s} \right) \right]^{s(\alpha_2 - \alpha_1)} + N_0(f),
    \label{eqn2}
\end{equation}
where \(s\) controls the smoothness of the transition between power-law regimes with indices \(\alpha_1\) and \(\alpha_2\). 
The noise floor, $N_0 (f)$, is modeled according to the DFB and SCM noise floors.
We set $\alpha_1 = -8/3$ and $\alpha_2 = -10/3$ for the density spectrum. For the magnetic spectrum, we set \(\alpha_2 = -16/3\), considering a KA to entropy cascade transition, and $f_b$ is set to $100\,\mathrm{Hz}$.
Fig.~\ref{fig5}(a,b) shows both the noise-free (dashed) and noise-added (solid) model spectra, using parameters that best describe our dataset: SNR $\simeq500$ (at $f=f_b$) for $n$; SNR $\simeq50$ for $B$ with $s = 0.2$.
Fits to the noise-added empirical spectra in the frequency range $f_b<f<f_{\mathrm{SNR=2.5}}$ and the corresponding slopes (in red) are indicated. 
Fig.~\ref{fig5}(c,d) shows the variation of the fitted slopes on the empirical spectrum of $n$ and $B$ with the smoothness parameter $(s)$ and SNR. 
The fits in the top panel shows that the slopes can become substantially shallower, with $-3.09$ for the $n$ spectrum and $-4.72$ for the $B$ spectrum with the available SNR in our dataset.
This accounts for $n$ spectral slopes being slightly shallower than $-10/3$. The model also explains why the $B$ spectrum does not fully reach the expected steep $-16/3$ slope.
Since the $B$ spectrum is steeper and also has a lower SNR than the $n$ spectrum for our intervals, they are more likely to be affected by noise.
In our analysis, a strong SNR dependence of the $B$ spectral slopes is indeed observed in the sub-electron range (not shown).
The bottom panel shows that for reliable measurement of such steep slopes a SNR $\gtrsim10^3$ is required. For low SNR ($\lesssim100$) the slopes are quite sensitive to noise, which may explain the broader spread of the distribution of slopes at $f>f_b$ in Fig.~\ref{fig3}.
This analysis is consistent with the interpretation of Huang \textit{et al.} \cite{Huang_2014}, where they attributed the observed steeper magnetic spectral slopes ($\alpha_2\simeq -5$) to higher SNR in their data compared to Sahraoui \textit{et al.} \cite{Sahraoui_2013} ($\alpha_2\simeq -4$) in the solar wind.
That said, even after accounting for the low SNR, the measured magnetic slopes are shallower than predicted by our empirical model.
This may be due to limited frequency resolution (only up to $k_\perp \rho_e \sim 3$), which could mean that the asymptotic $k_\perp \rho_e \gg 1$ regime required to observe pure entropy cascade fluctuations is not fully reached. At scales close to $\rho_e$, some undamped KA or inertial-KA fluctuations may persist, resulting in magnetic slopes shallower than expected.
Nevertheless, the $n$ spectrum showing a near $-10/3$ slope and the $B$ spectrum being comparatively steeper for $k_\perp \rho_e>1$, even with $B$ having lower SNR than $n$, serve as key evidence consistent with the predictions of electron entropy cascade.

%%%%%%%%%%%%%%%%%%% PARAGRAPH 12 %%%%%%%%%%%%%%%%%%%%

It should be noted that these interpretations require Taylor's hypothesis (TH) to be valid, i.e., the plasma-frame fluctuation frequency should be much smaller than the advection rate past the spacecraft ($\omega \ll |\mathbf{k} \cdot \mathbf{V}_{\rm rl}|$).
While $\omega$ is difficult to measure directly, it can be estimated as $\omega \sim \tau_\mathrm{nl}^{-1}$ assuming critical balance \cite{goldreich_sridhar_1995}, where $\tau_\mathrm{nl}$ is the nonlinear entropy cascade time \cite{Schekochihin_2008, Schekochihin_2009}.
The TH condition then reduces to $\delta n/n \ll (k_\perp\rho_e)^{-1/2} V_\mathrm{rl}/v_{\mathrm{th},e}$, where $v_{\mathrm{th},e}$ is the electron thermal velocity. For the typical conditions in our intervals, this implies that for TH to hold $\delta n / n \ll 0.1$ at $k_\perp \rho_e \sim 1$, which is well satisfied. 
}

\noindent
\paragraph*{Conclusion:}{

%%%%%%%%%%%%%%%%%%% PARAGRAPH 13 %%%%%%%%%%%%%%%%%%%%

We have presented high-resolution density spectra obtained from spacecraft potential measurements, alongside magnetic spectra resolving scales smaller than $\rho_e$, leading to the following main findings: At sub-$\rho_e$ scales, the density and magnetic spectra diverge, with the density spectrum exhibiting slopes close to $-10/3$ and the magnetic spectrum  comparatively steeper. 
These results indicate that turbulence at these scales becomes predominantly electrostatic and are consistent
with the theory of electron entropy cascade \cite{Schekochihin_2008, Schekochihin_2009}, a phase-space process, which can explain how electrons are irreversibly heated in a wide range of space and astrophysical systems.
In this context, our study motivates the need for improved particle instruments to directly measure the associated small-scale velocity-space structure and to fully understand the role of nonlinear phase-space dynamics in the irreversible dissipation of turbulent energy in weakly collisional plasmas.
We also note that, while the magnetic spectrum is steeper than the density spectrum, it appears shallower than predicted by the entropy cascade theory. This discrepancy may arise from instrumental noise, allowing the possibility that the true magnetic spectrum may satisfy a steeper power law than observed. Alternatively, the shallower magnetic spectrum may result from residual, weakly damped electromagnetic fluctuations near $\rho_e$. Future theoretical developments and improved measurements may help resolve this discrepancy, with our results providing important quantitative constraints.
}

%TC:ignore
\begin{acknowledgments}
S.M. is supported by a QMUL PhD studentship. C.H.K.C. is supported by UKRI Future Leaders Fellowship MR/W007657/1. C.H.K.C. and D.M. are supported by STFC Consolidated Grant ST/X000974/1. S.M. acknowledges A.A. Schekochihin, P.A. Simon, C. Sishtla, S. Greess and F. Koller for useful discussions. 
The authors acknowledge helpful discussions with D. Malaspina, M. Pulupa, M. Liu and J. Bonell regarding the spacecraft potential data.
Data used are publicly available at the PSP FIELDS (\href{https://fields.ssl.berkeley.edu/}{fields.ssl.berkeley.edu}) and SWEAP (\href{http://sweap.cfa.harvard.edu/}{sweap.cfa.harvard.edu}) data repositories.
\end{acknowledgments}
%TC:endignore

%\appendix

%Calibration of spacecraft potential measurements with the electron quasi-thermal noise density for the interval $2022-06-04,12-18$.

\bibliographystyle{apsrev4-2}
\bibliography{main}

\end{document}